# Power and the Pandemic: Exploring Global Changes in Electricity Demand During COVID-19

Elizabeth Buechler[a,+], Siobhan Powell[a,+], Tao Sun[b,+], Chad Zanocco[b,+], Nicolas Astier[c,+], Jose Bolorinos[b], June Flora[b,d], Hilary Boudet[e,*], Ram Rajagopal[b,*]

[a] Mechanical Engineering, Stanford University, Stanford, CA 94305 USA
[b] Civil and Environmental Engineering, Stanford University, Stanford, CA 94305 USA
[c] Economics, Stanford University, Stanford, CA 94305 USA
[d] Stanford Solution Science Lab, Stanford University, Stanford, CA 94305 USA
[e] School of Public Policy, Oregon State University, Corvallis, OR 97331 USA

[+] These authors contributed equally to this research

[*] Corresponding Authors:
    Hilary Boudet, School of Public Policy, Oregon State University, Hilary.Boudet@oregonstate.edu
    Ram Rajagopal, Civil and Environment Engineering, Stanford University, ramr@stanford.edu

**Abstract**
Understanding how efforts to limit exposure to COVID-19 have altered electricity demand provides insights not only into how dramatic restrictions shape electricity demand but also about future electricity use in a post-COVID-19 world. We develop a unified modeling framework to quantify and compare electricity usage changes in 58 countries and regions around the world from January-May 2020. We find that daily electricity demand declined as much as 10% in April 2020 compared to modelled demand, controlling for weather, seasonal and temporal effects, but with significant regional variation. Clustering techniques show that four impact groups capture systematic differences in timing and depth of electricity usage changes, ranging from a mild decline of 2% to an extreme decline of 26%. These groupings do not align with geography, with almost every continent having at least one country or region that experienced a dramatic reduction in demand and one that did not. Instead, we find that such changes relate to government restrictions and mobility. Government restrictions have a non-linear effect on demand that generally saturates at its most restrictive levels and sustains even as restrictions ease. Mobility changes, though more consistent across countries regardless of government restrictions, are also linked to demand changes. Steep declines in electricity usage are associated with workday hourly load patterns that resemble pre-COVID weekend usage. Quantifying these impacts using a unified modeling framework is a crucial first step in understanding the impacts of crises like the pandemic and the associated societal response on electricity demand.

**Significance Statement**
Understanding how efforts to slow the spread of COVID-19 may be linked to changes in electricity demand can provide insights about future electricity use during and after COVID-19. We develop a unified modeling framework, allowing us to quantify and compare electricity usage changes in 58 countries and regions around the world from January-May 2020. We also explore how such changes relate to government restrictions and mobility. We find that global electricity demand declined by as much as 10% in April compared to modelled demand, controlling for weather, seasonal and temporal effects, but with significant variation that can be characterized into four groups, ranging from mild change of 2% to an extreme decline of 26%.

**Abstract**

Understanding how efforts to limit exposure to COVID-19 have altered electricity demand provides insights not only into how dramatic restrictions shape electricity demand but also about future electricity use in a post-COVID-19 world. We develop a unified modeling framework to quantify and compare electricity usage changes in 58 countries and regions around the world from January-May 2020. We find that daily electricity demand declined as much as 10% in April 2020 compared to modelled demand, controlling for weather, seasonal and temporal effects, but with significant regional variation. Clustering techniques show that four impact groups capture systematic differences in timing and depth of electricity usage changes, ranging from a mild decline of 2% to an extreme decline of 26%. These groupings do not align with geography, with almost every continent having at least one country or region that experienced a dramatic reduction in demand and one that did not. Instead, we find that such changes relate to government restrictions and mobility. Government restrictions have a non-linear effect on demand that generally saturates at its most restrictive levels and sustains even as restrictions ease. Mobility changes, though more consistent across countries regardless of government restrictions, are also linked to demand changes. Steep declines in electricity usage are associated with workday hourly load patterns that resemble pre-COVID weekend usage. Quantifying these impacts using a unified modeling framework is a crucial first step in understanding the impacts of crises like the pandemic and the associated societal response on electricity demand.

**Significance Statement**

Understanding how efforts to slow the spread of COVID-19 may be linked to changes in electricity demand can provide insights about future electricity use during and after COVID-19. We develop a unified modeling framework, allowing us to quantify and compare electricity usage changes in 58 countries and regions around the world from January-May 2020. We also explore how such changes relate to government restrictions and mobility. We find that global electricity demand declined by as much as 10% in April compared to modelled demand, controlling for weather, seasonal and temporal effects, but with significant variation that can be characterized into four groups, ranging from mild change of 2% to an extreme decline of 26%.

**Main Text**

**Introduction**

Choices made to stop the spread of COVID-19 – by individuals, local entities and national governments – have the potential to profoundly impact electricity usage patterns. Moreover, changes taking place during the pandemic may persist long after the public health crisis subsides. We develop a unified modeling framework to quantify and compare electricity usage changes across countries and regions. This framework provides a crucial – and previously missing – first step in drawing meaningful insights about the pandemic's impact on electricity use. It also creates a valuable tool for researchers and other stakeholders in the electricity sector to quantify and compare changes in demand in the wake of past and future events on a global scale.

We apply this framework to electricity demand data from 58 countries and regions around the world, representing about 60% of the world's population and 75% of global electricity demand (1), from January-May 2020. For the 53 countries with daily data, we find that total electricity demand declined by as much as 10% in April 2020 compared to modelled demand, controlling for weather, seasonal and temporal effects. This dramatic decline is on par with some estimates of the effect of the 2008 global financial crisis (2). Yet, it masks significant variation among countries and regions in our sample – variation that is missed in studies that focus on specific regions or countries (3-10). Surprisingly, despite much diversity in the countries and regions studied, we are able to characterize this variation using four impact groups based on the depth and timing of decreases in electricity demand – extreme, severe, moderate and mild – but these groups do not align with traditional geographic distinctions (e.g., developed vs. developing; east vs. west; continents). Instead, we explore how government restrictions and mobility patterns relate to electricity demand changes, finding these linkages to be most pronounced in the extreme and severe impact groups and less so in the moderate and mild ones.

Crises like COVID-19 can serve as focusing events (11-13), creating windows of opportunity for change. Indeed, large-scale power outages have triggered significant policy and regulatory changes aimed at shoring up aging infrastructure and "smartening" the grid (14-16). Some energy analysts have called on world leaders to seize the "opportunity" created by the pandemic to implement lasting policy and structural changes to transform existing energy systems and address the climate crisis (17-19) – to which electricity generation from fossil fuels is a major contributor. Our results help to clarify our understanding of the link between COVID-19 pandemic-related social and economic restrictions and electricity demand. This, in turn, can provide important lessons going forward, not only helping utilities, companies that purchase power directly, and public officials to respond more effectively to the impacts of future viral outbreaks, but also ongoing efforts to build more resilient, sustainable infrastructure systems (20-22).

Existing analyses of COVID-19's impacts on electricity demand have been limited either by their geographic scope or an incomplete accounting of seasonal, weather and temporal effects (3-8, 10, 23-26). Recent global studies have been more focused on impacts to carbon emissions than electricity demand (17, 23, 27). More importantly, these studies have yet to develop a clear, open framework that allows for broad comparisons of changes in electricity demand worldwide. The International Energy Agency's recent report, *Global Energy Review 2020*, represents the most comprehensive effort to date to analyze COVID-19's global impact on energy use but is limited in scope for electricity (28). Our unified modeling framework learns accurate region-specific regression models for electricity consumption despite heterogeneity in load behavior. In contrast to previous studies (3, 5, 8, 23, 25, 26, 29), we also use extensive out-of-sample validation for both model selection and to verify the generalization capability of the learned models to new datasets.

As both a product and recipient of change, electricity demand does not take place in a vacuum. The Great Depression, World War II, the postwar period, and the 2008 global economic recession all had profound and lasting impacts on electricity use – and power systems – in the U.S. and around the world (30-32). There is every reason to believe that COVID-19 is an event of similar importance. Yet, we have few global, comparative studies of the effects of these and other types of events on electricity consumption. This study is designed not only to provide valuable insights into the relationship between the pandemic and electricity use on a global scale but also to provide a strong, enduring framework for future studies.

**Results and Discussion**

*Changes in Electricity Demand at Scale*

Our examination of changes in electricity demand across 58 countries and regions reveals significant variation in the impacts of the COVID-19 pandemic on electricity use (**Fig. 1**). Estimated daily electricity reductions in demand range from the extreme (e.g., India's March to May reductions average 15%) to little impact at all (e.g., Australia's 2% average daily drop over the same period). These declines roughly align with the timing and severity of government restrictions, or confinement index (CI) levels, as measured by Le Quere et al. (17) (*SI Appendix*, **Table S2**). The most drastic reductions in demand occur in April, when policies were most restrictive, except in China where restrictions were most severe in February. Impacts subside in May, as countries began to reopen.

Focusing on the U.S. and Europe specifically, European countries – particularly a group of highly impacted countries in Southern Europe (e.g., Italy, France, Spain, among others) – generally showed larger decreases in demand than U.S. regions, with the exception of Northern Europe (e.g., Sweden, Denmark, Finland) where little change occurred. In Asia, both China and India experienced substantial decreases when restrictions were put in place, while Japan experienced a more muted change. For countries in South America (Argentina and Brazil), North America (Mexico) and Africa (South Africa), we also see examples of steep reductions, alongside countries with little impact (e.g., Chile). Some of this variation could be due to more restrictive government policies and enforcement; cultural differences in acceptance and compliance with these policies; electricity demand composition (e.g., industrial, commercial, residential); or a combination of these and other factors. The result is a patchwork of electricity use impacts that vary widely in magnitude across continents and hemispheres with almost every continent having at least one country that experienced a dramatic reduction in demand and one that did not.

*Clusters of Electricity Demand Change*

Despite global diversity in the economy, policy, and culture in the countries and regions studied, we surprisingly discover that changes in electricity demand for the 53 countries and regions in our sample for which daily electricity demand is available cluster into four discrete groups, each distinguished by how early demand changes start and how deep they go (**Fig. 2a**).

The "Extreme" impact group, comprised of Italy and India, experienced the steepest change in weekly electricity demand (averaging 26% at their lowest point in late March). The "Severe" impact group includes the most countries and regions in our sample (N=23) – many European countries, two U.S. regions (Tennessee and North / South Carolina), New Zealand, Brazil, and Mexico – and exhibited average electricity use reductions of as much as 12%. The "Moderate" impact group had electricity use reductions averaging as much as 7%, generally occurring later in mid-April, and includes many U.S. and Canadian regions, several European countries (e.g., Denmark and Switzerland), Japan, Chile, Russia, and Singapore. We note that a 7% decrease is similar to some estimates of the impact on electricity demand during the 2008 global financial crisis (2), indicating that many of the electricity-demand reductions that we see in relation to the COVID-19 pandemic are unprecedented. Finally, the "Mild" impact group consists of Scandinavia, Latvia, Ukraine, Australia, and three U.S. regions (Northwest, Southwest and Florida). These areas experienced lesser, though still sizable, reductions in electricity demand, averaging 2% at its lowest point, generally occurring in late April, though some members of this group had no impacts or even increases during the period.

Even more surprisingly, countries with very diverse economies and cultures group together, reinforcing our finding that global geography has a limited role in determining COVID-19's grid impacts. Instead, government restrictions and how the population responds to them could be key indicators of demand change, which we turn to now.

*Relationship to Government Restrictions*

One factor that is likely to account for differences in electricity use responses to COVID-19 is the severity of government restrictions introduced during the pandemic. To analyze the role of COVID-19 restrictions, we used the confinement index (CI) developed by Le Quere et al. (17), which orders restrictions into three (increasing) levels of severity (*SI Appendix*, **Table S2**). Because most countries in our sample moved progressively in and out of CI levels (from CI 1 to CI 2 to CI 3 and back out to CI 2 and then CI 1) during the period under examination, we further delineated CI 1 and CI 2 periods as early (pre-CI 3) and late (post-CI 3).

Bivariate regressions of each country's or region's 2020 daily electricity use change and its time-varying CI level (N=50 countries and regions with both datasets available) show that the relationship between government policies (CI levels) and electricity demand is generally larger in countries and regions in the Extreme and Severe groups (e.g., average CI level 3 decreases of 18% and 10%, respectively) than those in the Moderate and Mild groups (e.g., average CI level 3 decreases of 5% and 2%, respectively) (**Fig. 2b;** *SI Appendix*, **Table S3a)**. We also find that, the magnitudes of the decrease during CI 1 and CI 2 periods are smaller in earlier than later periods, as the impact of the peak restrictions appears to continue into the later periods.

In addition to overall impact of CI level captured by these indicators, we used multi-adaptive regression splines (MARS) techniques (33) to examine the effect of each additional day in a particular CI level on electricity demand. The model fit improves on the simple CI regressions, with $R^2$ values ranging from 0.25 to 0.91 **(***SI Appendix*, **Table S3b).** For most countries and regions, the impact of government restrictions achieves a saturation effect after several days at each level beyond which increasing days at that CI level result in little change in demand. If saturation is defined as a non-significant deviation from zero of per day impact, 86% of countries and regions experience saturation during at least one CI level; and, of the countries and region that spend time in CI 3, more than 48% achieve saturation during this period (*SI Appendix*, **Fig. S4a).**

Prior to saturation, the per day reduction in demand has different patterns for each impact group. In the Extreme group, the per day impact of CI 3 (1.24) is higher than that of the early CI 2 period (0.84). In the Severe cluster, this is reversed: the per day impact of the early CI 2 period (0.87) is higher than CI 3 (0.41). In the Moderate and Mild group, the per day impact is similar for both early CI 2 (0.25 and 0.08, respectively) and CI 3 (0.32 and 0, respectively) **(***SI Appendix*, **Fig. S4b).** The breakpoints (or knots) identified by the model also reveal delays between the transition to a new CI level and a change in demand: many regions enter the early CI 1 period two to four weeks before demand drops. In contrast, CI 3 and early CI 2 restriction periods are the most likely to have an immediate impact (*SI Appendix*, **Fig. S4c)**. Moreover, recovery is limited through the late CI 2 and CI 1 periods, particularly in the Severe and Moderate impact clusters **(***SI Appendix*, **Fig. S4d).** Taken together, our findings indicate that CI 2 restrictions are linked to the steepest declines in electricity demand in most countries and regions, and CI 3 appears to have the most impact in countries in the Extreme group, where enforcement of restrictions was likely more intense **(***SI Appendix*, **Fig. S4b).**

Yet, variation in electricity demand is not fully explained by government restrictions. For example, the timelines of CI restriction levels in Australia and New Zealand are very similar, but their changes in electricity demand are very different, placing them in two different impact groups: Mild and Severe, respectively. To better understand these differences, we explore changes in daily travel patterns or human mobility.

*Relationship to Changes in Mobility*

Government restriction levels do not measure variations in compliance, enforcement, or the true impact of these restrictions on human behavior. Changes in mobility patterns (34) can serve as one proxy for these differences.

Overlaying changes in mobility, electricity use, and government restriction levels for each of our countries and regions arranged by electricity impact grouping (**Fig. 3a;** *SI Appendix*, Fig. S5), two key insights emerge. First, all regions show roughly similar changes in some types of mobility (~50% decreases in workplace mobility; 20-30% increases in residential mobility), regardless of CI intensity or electricity use change. Such declines are evident even in places like Florida with no visible electricity use change. Second, while mobility declines usually coincide with electricity use reductions, there are some exceptions. In California and New York, for example, workplace mobility starts to decrease in early March, one to two weeks before government restrictions reach CI 3 levels or electricity use declines.

These findings suggest a complex and heterogeneous relationship between mobility and electricity use, potentially linked to changes in workplace policies prior to government restrictions, individual decisions to limit mobility as a precautionary measure, or other grid-related factors, such as the presence of distributed generation and/or electricity demand composition.

Estimating the demand elasticity of each country's or region's 2020 daily electricity use change with respect to its daily mobility change in each category, shows that increases in residential mobility (i.e., staying home more, an indication of more effective confinement policies) are most directly related to decreases in electricity demand, while increases in all other types of mobility are linked to increases in electricity demand (**Fig. 3b;** *SI Appendix*, **Fig. S5**). Decreases in workplace mobility are the next most tightly linked to demand reductions, though several other mobility categories (e.g., grocery/pharmacy, transit) seem to have quite similar effects. Our electricity impact groups also align quite well with mobility changes: the relationship between electricity and mobility change is most pronounced for the Extreme group and least pronounced for the Mild group.

$R^2$ values for country models (**Fig. 3c**; *SI Appendix*, **Table S2b**) show that – even if changes in mobility patterns were roughly similar across most of the countries and regions in our sample – small differences in day-to-day mobility changes are able to capture finer grained electricity use changes than CI levels. The link between mobility and demand is also stronger in Extreme and Severe groups. This finding suggests that mobility itself may be a suitable proxy for a range of human responses to the COVID-19 pandemic, some of which may not be related to the level of government-imposed restrictions.

*Changes in Daily Electricity Demand Patterns*

For some countries/regions, we observe a change in load shape patterns (hourly electricity demand over a day)[1] that has been noted in the media (24): workday load shapes in April 2020 resemble weekend load shapes in previous years (**Fig. 4**). The extent to which this observation matches observed load patterns, however, varies by impact group **(Fig. 4**; *SI Appendix*, **Fig. S6)**. The workday-to-weekend shift prevails in the Extreme and Severe groups. The other groups are more heterogenous: some countries and regions in the Moderate impact group exhibit this shift; most countries in the Mild impact group do not. These differences could be due to more extensive limits on industrial and commercial operations in the more impacted groups, leaving residential users to drive demand on workdays.

Such deep changes in load patterns can have a major impact on baseload and peak demand, respectively defined as the minimum and maximum hourly load levels on a given day. For example, low baseload levels can trigger curtailments in excess generation and make it more difficult to meet peak demand (29).

Comparing workdays in April 2020 to those in previous years (April 2016 to April 2019), we find that the average drop in baseload and peak demand aligns well with our impact groups, ranging from little to no average impact in the Mild group to average decreases in baseload and peak of 3% and 6%, respectively, for the Moderate group, 10% and 12% for the Severe group, and 17% and 20% for the Extreme group (without correcting for temperature differences). In terms of timing, peak demand was more affected than baseload. Particularly in countries and regions where electricity demand previously peaked in the morning, it shifted later in the day, either the late morning (e.g., France, Germany, Spain) or the evening (e.g., Italy, Belgium, Ireland, New-Zealand, United Kingdom).

These dramatic changes in load shapes also affected the ability of grid operators to predict future demand. Comparing week-ahead forecasts to actual demand where this data is available (27 European countries and California), we find many over-predictions during this period, which took approximately 3-5 weeks to correct (*SI Appendix*, **Fig. S7**).

**Conclusions**

COVID-19's impact on electricity demand in some places (i.e., Italy and India) has been unprecedented, but not everywhere. Instead, responses can be grouped into four categories, with maximum reductions ranging from 26% (Extreme Impact) to 2% (Mild Impact). Variation in these responses within global regions is often greater than between them, even if notable regional differences remain (e.g., impacts in Europe were generally larger than in the U.S.).

Links between government restriction levels and electricity demand change are most pronounced in the Extreme and Severe impact groups and less so in the Moderate and Mild impact groups. Such variation is likely due to many factors, including differences in sectoral loads, behavioral choices in terms of compliance with restrictions, and perhaps even COVID-related deaths and media coverage of the pandemic (which could change behavior absent restrictions). Government restrictions generally have a non-linear effect on demand that begins slowly, accelerates as restrictions tighten but then generally saturates at its most restrictive levels and sustains even as restrictions ease. CI level 2 restrictions are key in demand reductions, suggesting that mandatory closure of educational institutions, public buildings, and non-essential businesses is largely responsible for demand decreases. Mobility offers a sharper focus on electricity demand change with workplace and residential mobility strongly linked at the daily level. In fact, mobility could be utilized to anticipate demand changes for grid operations and planning. Analysis of hourly load patterns and reductions in average daily load levels and baseloads shows that many system operators have faced unprecedented changes and have had to make significant adjustments in operations and planning.

---

[1] Hourly consumption data is generally published net of distributed generation, so observed load shape changes may not always be driven by COVID-19 restrictions, especially in regions and countries that have recently experienced increases in solar PV generation (e.g. California, Australia).

This work represents an ongoing effort to characterize the pandemic's impacts on electricity use. Understanding changes in electricity demand in the early stages of the COVID-19 pandemic can provide valuable insights into how to respond to critical challenges energy systems are likely to face in the future. As the pandemic reaches new stages on a path toward global recovery, energy systems will continue to face an uncertain future. They will undoubtedly be struck by other crises driven, for example, by extreme weather events or dramatic shifts in the economy. Our unified modelling framework provides a valuable tool for quantifying and comparing the impact of these events now and in the future. These estimated regional responses serve as a predictor of future impacts due to changes in confinement policies as the pandemic progresses or as a guideline for events that result in reduced consumption activity.

Data availability remains a significant challenge to complete a global picture, and we will continue to add geographic regions. We also plan to expand our analysis in multiple, policy-relevant ways: incorporating sectoral level (i.e., residential, commercial, industrial) impacts, comparing recoveries across countries, further examining impact group and policy relationships, and assessing impacts on sources and carbon emissions from electricity generation.

The long-term influence of the pandemic on electricity demand and our energy systems remains uncertain. COVID-19 is likely to have profound societal and economic impacts that continue to reshape the global community in untold ways. Electricity demand and access are likely to experience significant changes in a post-COVID-19 world (35, 36). Whether policy makers will seize the "opportunity" created by the crisis to reimagine and reinvent energy and power systems remains an open question. While no one can anticipate when the next global crisis will strike or what it may bring, one critical crisis is lurking in plain sight: climate change. With its intricate connections to energy systems, understanding changes in electricity use in response to the COVID-19 pandemic could provide valuable lessons in our climate-changed world.

**Methods**

*Overall Approach*

We developed a unified modeling framework to explore relationships between changes in electricity demand, mobility, and government restrictions (*SI Appendix*, **Fig. S8**). We first estimated region-specific regression models to predict electricity consumption without the pandemic, accounting for weather, seasonal and temporal effects. Then, we calculated changes in daily electricity demand by comparing the baseline demand estimates from our learned regression models with actual demand data. We used cluster analysis to uncover common patterns in demand change over time across countries and regions. We then regressed the estimated daily electricity demand change on indicators of government restrictions and mobility changes to explore relationships between these factors. Finally, we examined changes in hourly load shape patterns and week-ahead forecasts.

*Demand Change Estimation*

The effects of different factors on electricity consumption vary substantially by region. This heterogeneity depends on climate, sectoral load composition, and various other societal factors. Our unified modeling framework learns accurate region-specific regression models for electricity consumption despite this heterogeneity. The regression models account for weather, seasonal, and temporal effects by using region-specific data for electricity consumption, population-weighted weather data, and a holidays database. We also used out-of-sample validation for both model selection and to verify the generalization capability of the learned models to new datasets.

Model parameters were identified using ordinary least squares (OLS) regression. The polynomial order of weather terms for each country was selected using k-fold cross validation with 10-folds. Higher order terms account for non-linear temperature dependence observed in the data but neglected in other studies (5, 23, 25, 26, 29). Validation errors were defined in terms of the root mean squared error (RMSE) of the predictions, normalized by the average demand for each region. The final regression model was fit using the combined dataset from all folds. Holidays were manually included in the model based on observations of the sensitivity of demand to each holiday in historical data. Daily out-of-sample validation errors for each country ranged between 1.27-4.91% RMSE. Model errors over longer time periods (e.g., weekly, monthly) are substantially lower (*SI Appendix*, **Demand Change Estimation)**.

*Cluster Analysis*

We applied K-Means clustering to the weekly electricity demand reductions to investigate similarities in the response between different countries. We utilized the elbow curve of cluster inertia to select K = 4 clusters (*SI Appendix*, **Fig. S2**).

*Modeling the Relationship between Electricity and CI/Mobility*

We used our calculations of daily electricity use change and OLS regression to model the relationship between changes in electricity, CI level, and mobility. The Multivariate Adaptive Regression Spline (MARS) model was fit using the cumulative number of days in each CI level to estimate the additional impact of time spent at each level.

*Statistical Analysis of Forecasting Accuracy*

Week-ahead forecast and actual demand data from 27 European countries and California was used to measure how well system operators predicted electricity demand. We used t-tests to test for the significance of the difference between week-ahead forecasted and actual demand in each country, region and week.

**Acknowledgments**

We thank Arun Majumdar and Gustavo Cezar for their valuable insights about this research. We are grateful to the organizations that provided open access to data about electricity demand, mobility, and weather, without which this research would not have been possible. Funding was provided by the National Science Foundation through the Smart & Connected Communities program (#1737565), the CONVERGE COVID-19 Working Groups for Public Health and Social Sciences Research, a CAREER award (#1554178) and a Graduate Research Fellowship, as well as a Stanford Graduate Fellowship.

26. McWilliams B & Zachmann G (2020) Electricity Consumption as a Near Real-time Indicator of COVID-19 Economic Effects.  (IAEE Energy Forum, Covid-19 Issue 2020).
27. Declercq B, Delarue E, & D'haeseleer W (2011) Impact of the economic recession on the European power sector's CO2 emissions. *Energy Policy* 39(3):1677-1686.
28. International Energy Agency (2020) Global Energy Review 2020: The impacts of the Covid-19 crisis on global energy demand and CO2 emissions.
29. Graf C, Qualgia F, & Wolak FA (2020) Learning about Electricity Market Performance with a Large Share of Renewables from the COVID-19 Lock-Down.
30. Bakke G (2016) *The Grid: the fraying wires between Americans and our energy future* (Bloomsbury Publishing USA).
31. Carley S, Nicholson-Crotty S, & Fisher EJ (2015) Capacity, Guidance, and the Implementation of the American Recovery and Reinvestment Act. *Public Administration Review* 75(1):113-125.
32. Hecht G (2009) *The radiance of France: Nuclear power and national identity after World War II* (MIT press).
33. Milborrow S & Derived from mda:mars by T. Hastie and R. Tibshirani (2018) earth: Multivariate Adaptive Regression Splines).
34. Google (2020) COVID-19 Community Mobility Reports.
35. Graff M & Carley S (2020) COVID-19 assistance needs to target energy insecurity. *Nature Energy* 5(5):352-354.
36. Castán Broto V & Kirshner J (2020) Energy access is needed to maintain health during pandemics. *Nature Energy* 5(6):419-421.
10

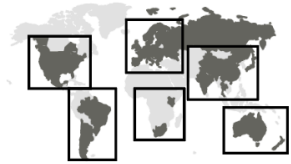
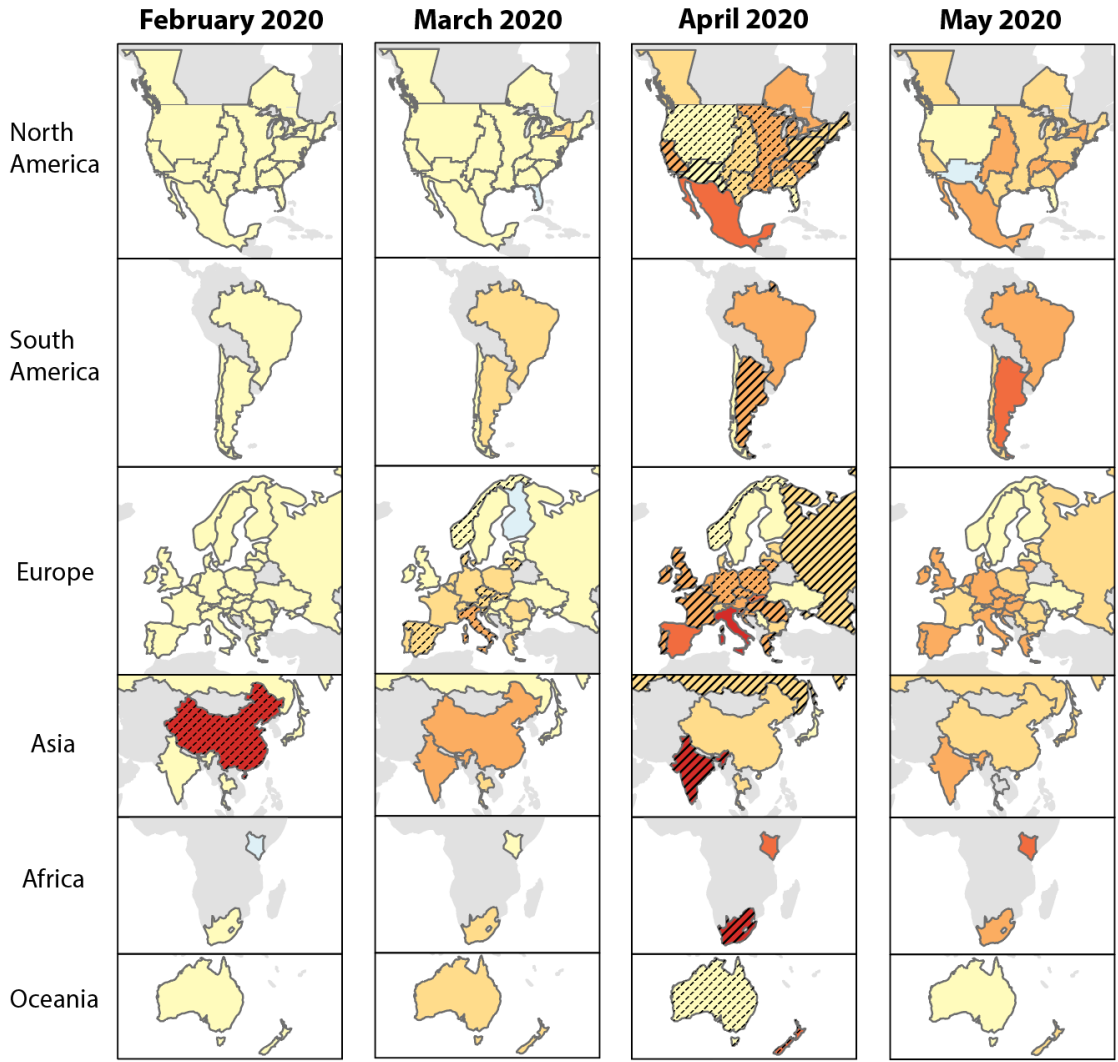
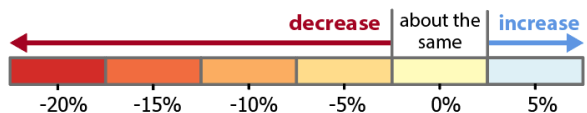
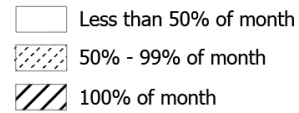

**Fig. 1. Estimated difference in modeled and actual electricity demand for countries and regions across the world, February - May 2020**. Color intensity corresponds to change from modeled electricity demand (+/- 2.5% per labeled midpoint in legend) and hatched fill the number of days under Confinement Index (CI) Level 3 (i.e., national policies that substantially restrict the daily routine of all but key workers) in each month. Gray areas indicate missing data.

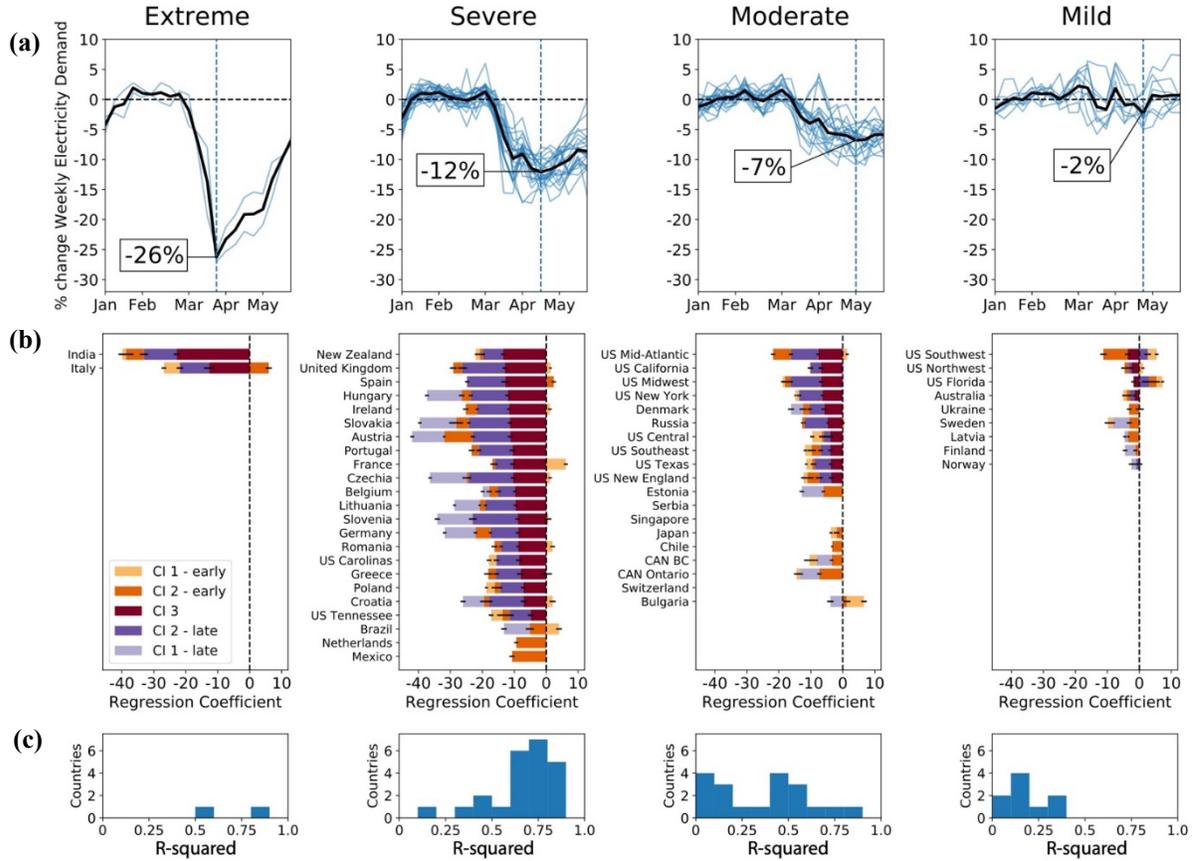

**Fig. 2. Electricity demand impact groups during COVID-19 and relationships to government restriction levels**
**(a)** Clustering countries and regions by the percent change in modeled and actual weekly electricity demand, with four clusters, Jan 1 - May 31, 2020. Black lines represent mean change in electricity use per cluster; light blue lines are each country or region within the cluster. Vertical dashed lines indicate when the cluster mean reached its minimum, with the value noted.
**(b)** Beta coefficients for CI 1 early/late, CI 2 early/late, and CI 3 from ordinary least squares regression models of daily electricity change and CI level for each country/region from February 15 - May 31, 2020. No bar indicates the country/region spent zero days in the given CI level. All betas are significant at p<0.05 except Norway for CI 3; Norway, US Northwest, US Central, and US Florida for CI 2 late; Australia, Bulgaria, Croatia, Czechia, Denmark, France, Greece, New Zealand, Poland, Slovakia, Slovenia, US Midwest, US New England, US Northwest, US Central, US Tennessee and US Texas for CI 2 early; Norway and Ukraine for CI 1 late; and CAN British Columbia, CAN Ontario, Greece, India, Sweden, Russia, US Midwest, and US California for CI 1 early. Country-specific models are provided in the Supplementary Information.
**(c)** Distributions of $R^2$ values of the CI level regressions for the countries in each cluster. $R^2$ values suggest that this modeling approach provides a relatively strong model fit for the Extreme and Severe impact groups ($R^2$ values range from 0.58-0.82 and 0.19-0.85, respectively) with weaker fit for the Moderate and Mild groups ($R^2$ values range from 0.04 to 0.81 and 0.05 to 0.39, respectively) (*SI Appendix*, **Table S3a**).

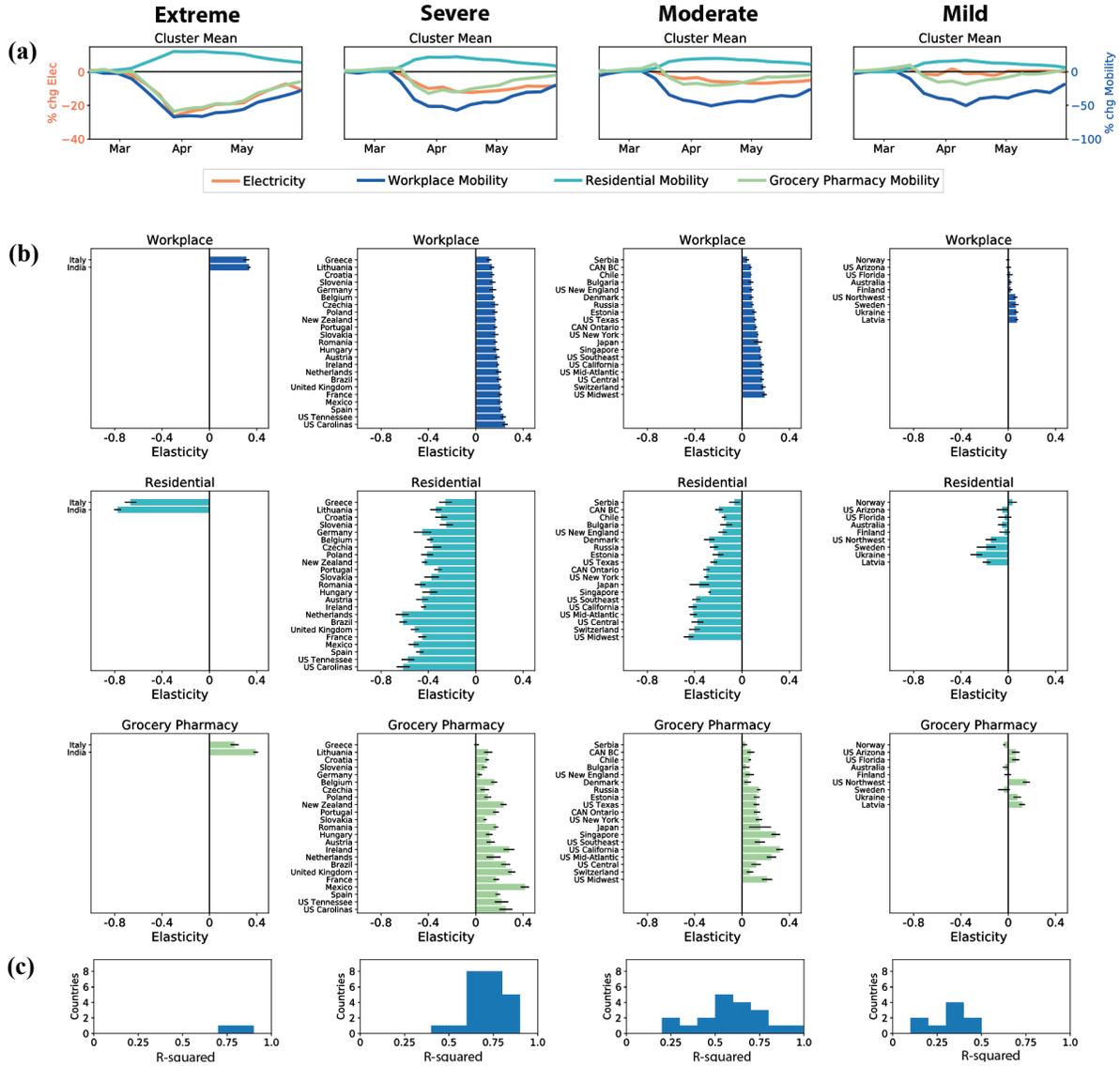

**Fig. 3. Electricity demand change impact groups during COVID-19 and relationship to changes in mobility (a)** Percent change in mobility and electricity demand across selected countries and regions by demand impact group. Orange lines represent cluster means for weekly average percent changes between forecasted and actual demand; other colored lines represent mean mobility changes for each location type.

**(b)** Elasticity coefficients measuring the relationship between changes in workplace, residential, and grocery/pharmacy mobility and changes electricity use. Elasticities for all metrics are included in the Supplementary Information (*SI Appendix*, **Fig. S5**). Error bars show the standard error for estimated elasticity coefficients. All are statistically significant ($p<0.05$) except Australia, Finland, US Florida, Norway, and US Southwest for workplace mobility; Finland, Australia, Serbia, US Florida, Norway, and US Southwest for residential mobility; and Greece, Germany, Japan, Bulgaria, Denmark, US New England, Serbia, Sweden, Finland, and Australia for grocery/pharmacy mobility. Country-specific models are provided in the Supplementary Information.

**(c)** Distribution of R-squared values for regression models predicting electricity demand using all six mobility metrics as regressors for each country and region in the cluster.

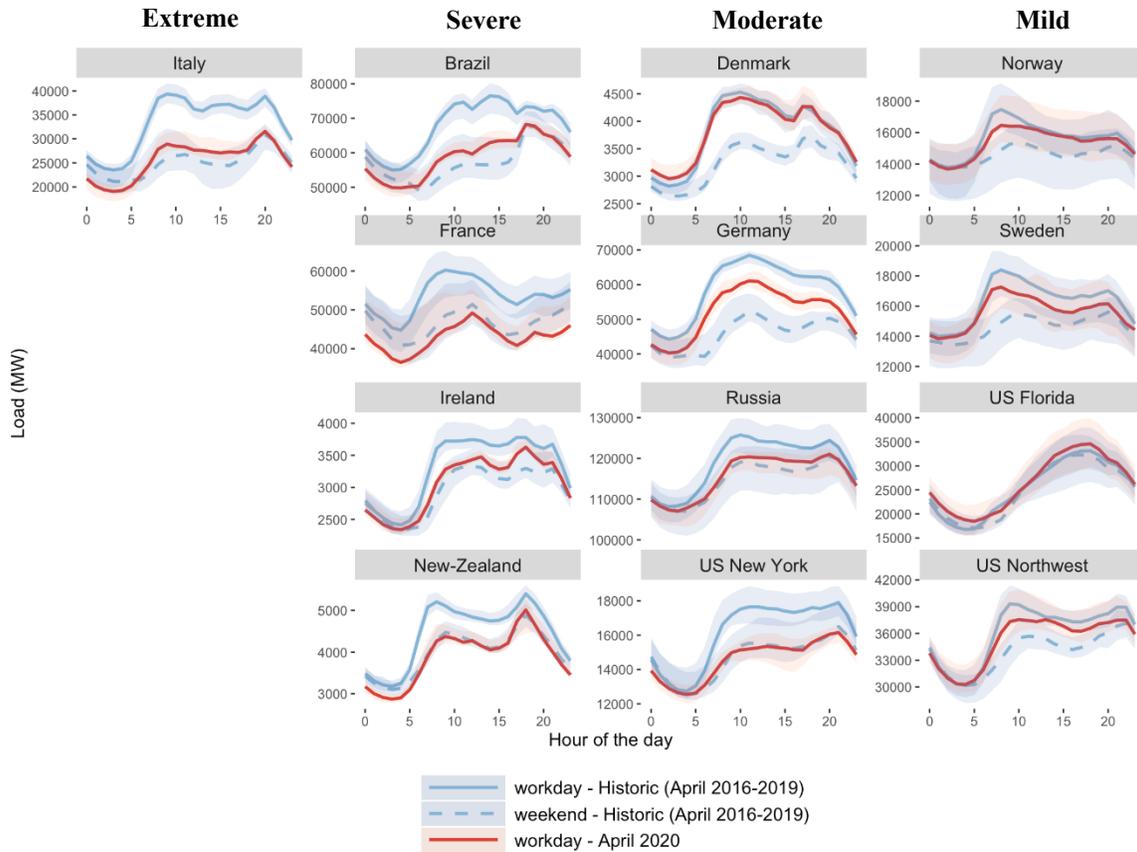

**Fig. 4. Comparison between April 2020 daily load curves and historic load curve patterns.** Range of observed daily load shapes for a sample of countries and regions (see Supplementary Information for other countries and regions), aligned by impact group. Blue depicts historic observations during the months April 2016, 2017, 2018 and 2019; red depicts observations in April 2020. Solid lines represent workdays; dashed lines weekends (holidays excluded). The plotted line corresponds to the median load level observed for this hour over the period of interest. Sleeves show the range between the 10th and 90th quantiles. Load levels have not been corrected for differences in temperature.

# Supplemental Information for:

# Power and the Pandemic: Exploring Global Changes in Electricity Demand During COVID-19

Elizabeth Buechler[a,+], Siobhan Powell[a,+], Tao Sun[b,+], Chad Zanocco[b,+], Nicolas Astier[c,+], Jose Bolorinos[b], June Flora[b,d], Hilary Boudet[e,*], Ram Rajagopal[b,*]

[a] Mechanical Engineering, Stanford University, Stanford, CA 94305 USA
[b] Civil and Environmental Engineering, Stanford University, Stanford, CA 94305 USA
[c] Economics, Stanford University, Stanford, CA 94305 USA
[d] Stanford Solution Science Lab, Stanford University, Stanford, CA 94305 USA
[e] School of Public Policy, Oregon State University, Corvallis, OR 97331 USA

[+] These authors contributed equally to this research

*Corresponding Authors:
    Hilary Boudet, School of Public Policy, Oregon State University, Hilary.Boudet@oregonstate.edu
    Ram Rajagopal, Civil and Environment Engineering, Stanford University, ramr@stanford.edu

**Extended Data and Methods**

**Data Sources.** Hourly electricity consumption for the United States was obtained for 13 different subregions [D1], which each represent an aggregation of multiple grid balancing authorities defined by the US EIA (Energy Information Agency). Hourly or sub-hourly electricity data was obtained at the country-level for the majority of European countries [D2]. We excluded Cyprus, Montenegro, Luxembourg, and Bosnia from the analysis due to a combination of a high percentage of anomalous values in the demand data and high model validation errors, leaving us with 29 countries. The hourly data for Australia [D3] represents the electricity consumption for five states (New South Wales, Queensland, South Australia, Tasmania, and Victoria) which make up the majority (~90%) of the country's population. Hourly data was obtained for two Canadian regions: Ontario [D4] and British Columbia [D5]. Data for Japan (hourly) is limited to the TEPCO utility [D6], which serves 35% of the population.[1] Data for New Zealand [D7], Russia [D8], Mexico [D9], Brazil [D10], and Chile [D11] are reported at the country-level with hourly resolution. Electricity data for India is reported at the country-level with daily resolution [D12]. Data for Singapore is reported at 30-minute intervals [D13]. Mean daily demand was calculated from all of these datasets for use in the regression model. Additionally, there are five countries included in our study where only monthly electricity consumption data was available from public sources: China [D14], South Africa [D15], Thailand [D16], Argentina [D17], and Kenya [D18].

Population-weighted daily heating degree day (HDD) and cooling degree day (CDD) data for US states and census regions were applied in the models for the US regions [D19]. Population-weighted HDD and CDD statistics were generated for all remaining regions using temperature data from weather stations in the ASOS Network [D20] (located in cities with more than 100,000 inhabitants) and for which measurements were available over the full 2015-2020 period.

We obtained a database of holiday dates for each country [D21], and then manually selected which holidays were included in the model based on observations of the sensitivity of demand to each holiday in the historical data. Dummy variables were also added for specific days of the year that typically experience reduced demand but are not official national or religious holidays, such as the days between Christmas and New Year's Eve and days in the first week of the year.

For metrics of human mobility patterns we used daily data published by Google in their COVID-19 Community Mobility Report [D22]. This dataset includes changes in mobility related to workplaces, transit stations, grocery stores and pharmacies, residences, parks, and retail or recreational activities. The Google data is made available at the state- or country-level, with a limited number of missing days which were removed by Google to preserve anonymity. Each metric is given as a percent change over a baseline from January 3rd to February 6th, 2020, accounting for the day of the week.

For measures of the severity of COVID-19 restrictions we use the confinement index (CI) dataset developed by [D23]. The CI levels range from 0 to 3: CI 0 for no restrictions, CI 1 includes travel restrictions and bans on mass gatherings, CI 2 includes measures such as school closures and border closures, and CI 3 is a lock-down. Further details are included in **Table S2.**

**Demand Change Estimation.** To estimate changes in electricity demand, we developed region-specific regression models for predicting typical electricity consumption. Demand reductions were calculated on a daily basis, by comparing the model estimate with actual demand from January to May 2020. The regression model for predicting daily mean demand is given by:

Equation 1. Regression model predicting daily average demand

$$d = \alpha_0 + \sum_{i=1}^{n_w} \alpha_{i,h} h^i + \sum_{i=1}^{n_w} \alpha_{i,c} c^i + \sum_{i=1}^{11} \alpha_{i,m} m_i \sum_{i=1}^{6} \alpha_{i,w} w_i + \sum_{i=1}^{n_y-1} \alpha_{i,y} y_i + \sum_{i=1}^{n_s} \alpha_{i,s} s_i$$

---
[1] https://www.tepco.co.jp/en/corpinfo/illustrated/business/business-scale-area-e.html

where $d$ is the mean daily demand, $h$ is the number of daily heating degree days (HDD), $c$ is the number of daily cooling degree days (CDD), and $m_i$, $y_i$, $w_i$, and $s_i$ are dummy variables for the month of the year, year, day of the week, and holidays, respectively. HDD and CDD values were calculated using a base temperature of 65°F. $n_y$ is equal to the number of years of data included in the training, validation and test sets and $n_s$ is the number of different holidays included in the model. The last month of the year, day of the week, and year in the dataset were used as a baseline, which is why the corresponding dummy variables are omitted in the model. The yearly dummy variables account for long-term, consistent load growth patterns and the monthly dummy variables account for seasonal effects not captured by the HDD and CDD terms. The holiday dummy variables account for demand reductions typically observed on national and religious holidays. The datasets used for training daily models include region-specific data for electricity consumption, weather, and holidays. The temporal resolution of the raw electricity consumption data ranges between 15-minute and daily granularity for different countries.

The parameters of the regression model were identified using ordinary least squares regression. The order of the polynomial HDD and CDD terms $n_w$ (between 1 and 4) was selected based on the out-of-sample validation error using k-fold cross validation with 10 folds. For most regions, the 10 folds were randomly selected from data between January 2016 and the end of February 2020. For some regions, data was only available beginning in January 2017 or January 2018. Therefore, at least two years of data was used for training and validation for each region. Validation errors for each fold were defined in terms of the root mean squared error (RMSE) of the predictions, expressed as a percent of the average demand in the dataset. The average validation error over all folds of the 10-fold cross-validation was used for model selection. For most regions, $n_w = 4$ was selected. The final regression model was fit using the combined dataset from all 10 folds.

Mean validation errors for each country range between 1.27-4.91% RMSE. In-sample training errors and out-of-sample validation errors for each individual country are shown in **Fig. S1**. Note that errors over longer time periods than one day, such as over a week or month, are substantially smaller than the daily errors shown. Daily training and validation errors are similar for each region, indicating that negligible model overfitting occurred.

Changes to electricity demand during the time when COVID restrictions were in place were estimated for each country by comparing the baseline demand estimates from the learned regression model with actual demand data. Confidence intervals for the estimated change in demand were estimated using the prediction intervals from the regression model.

The regression model defined in Equation 1 describes the relationship between daily demand and temperature variation, day of the week, seasonal factors, load growth, and holiday effects. These are the primary factors that affect demand for all regions and countries considered. However, the individual effect of each of these factors on demand varies considerably by region around the world. For example, the effect of holidays on the magnitude of electricity demand is generally much more pronounced in Europe than in the US. The dependence of demand on weather variables is much stronger for regions with more extreme climates, such as Russia. Our empirical results show that our unified modeling framework can accurately predict baseline demand despite heterogenous demand behaviors of different regions.

There are additional variables, such as economic factors, that may affect demand in specific regions, but not to a large enough degree to significantly affect the conclusions drawn in this study. Preliminary analysis indicates that the electricity demand for India may have some correlation with country-level exports and other economic variables. Additionally, demand for regions with irregular demand growth rates may not be fully captured using yearly fixed effects. Some regions have increasing behind-the-meter distributed PV generation that is incorporated in the electricity demand data, and not explicitly accounted for in the modeling framework. Certain small effects observed in our results, such as the slightly increasing demand trend for the US Southwest region during the end of our COVID-19 analysis period, may be partially attributed to some of these factors which were not accounted for in the modeling framework.

For five of the countries included in our study, only monthly electricity consumption data was available from public sources (China, South Africa, Thailand, Argentina, and Kenya). For these countries, our estimates rely solely on monthly observations from either January 2012 (South Africa, Thailand), 2015 (China) or 2017 (Argentina, Kenya) to December 2019. We fit a linear regression model with monthly dummy variables and make country-specific and parsimonious model adjustments based on observed historical patterns. First, for countries that experience substantial growth in their electricity consumption (China, Kenya, Thailand), we add a time trend linear in the number of months since the beginning of our dataset. Second, if a correlation between temperature and electricity consumption is observed in the dataset, we include either monthly CDD (Thailand) or both monthly CDD and HDD (Argentina, China and South Africa) regressors in the model. Finally, for China we include a dummy variable for the month of the year during which the Spring Festival occurred (either January or February) which was found to improve model performance.

As a result, monthly model estimates for the 2020 monthly electricity consumption are based on the average consumption during the corresponding month over the past few years, with some corrections for temperature and load growth with historical load patterns, suggest such corrections are necessary. These estimates are not as accurate as our estimates derived from more granular load data. Confidence intervals typically range from +/- 3% (South Africa, Thailand), +/-5% (Argentina, China) or +/-6% (Kenya).

**Modeling the Relationship between Electricity and Government Restrictions or Changes in Mobility.** Data with daily intervals were used to model the relationship between changes in electricity, confinement index (CI) level, and mobility (see Data Sources section). For the electricity data in the regressions, we used the daily outputs of the model for change in electricity demand for each country/region. For the US electricity regions which include counties from multiple states, population weighted averages were used to calculate regional series from the state-level CI and mobility data.

Ordinary least squares (OLS) regression was used to model the relationship between CI or mobility and electricity change:

Equation 2. Regression model predicting electricity demand using CI level
$$\Delta d(t) = \alpha_{11}\delta(CI(t) = CI_{11}) + \alpha_{21}\delta(CI(t) = CI_{12}) + \alpha_3 \delta(CI(t) = CI_3) + \alpha_{22}\delta(CI(t) = CI_{22}) + \alpha_{12}\delta(CI(t) = CI_{12}) + \beta$$

Equation 3. Elasticity of electricity demand against metrics of mobility
$$\Delta d = \alpha_{M_i} M_i + \beta$$

Equation 4. Regression of electricity demand against all metrics of mobility
$$\Delta d = \sum_i \alpha_{M_i} M_i + \beta$$

Where $t$ represents time in days and $\beta$ represents the intercept coefficient. The dependent variable, $\Delta d$, is the percent change in daily electricity demand. The independent variables are a mobility metric, $M_i$, where $i$ indexes the six different mobility metrics, or indicators for whether the country/region was under CI level 1, 2, or 3 on that day. We differentiate between tightening and loosening periods as early/late, indicating whether the period occurred before or after the peak confinement level. We denote CI 1 early as $CI_{11}$, CI 2 early as $CI_{21}$, CI 3 as $CI_3$, CI 2 late as $CI_{22}$, and CI 1 late as $CI_{12}$. The coefficients in the regression, $\alpha$, are indexed similarly.

Equation 3 gives the calculation of elasticities with mobility, $\alpha_{M_i}$, as presented in **Fig. S5**. Equation 4 regresses electricity demand against all six mobility metrics together. These results are included in the Supplementary Information for each country.

In the CI regressions, for regions where CI level 0 was included in the dataset, CI level 0 was taken as the baseline. For regions where the timeframe included no dates at CI level 0, CI level 1 was used as the baseline and the constant was taken to represent its effect over CI level 0.

Multivariate adaptive regression splines (MARS) was used to model the relationship with time spent in each CI level. A cumulative series was created for each CI level:

Equation 5. Cumulative measure of CI level

$$S_{ij}(t) = \sum_{\tau=1}^{t} \delta(CI(\tau) = CI_{ij})$$

for $ij$ in the set $T = \{11, 12, 21, 22, 3\}$ to indicate the CI level as in Equation 2. Using the cumulative series, $S_{ij}$, as independent variables, the MARS model fit the electricity demand using a set of linear and linear-hinge functions:

Equation 6. MARS model predicting electricity demand using CI level cumulative metrics

$$\Delta d(t) = \sum_{ij \in T} \alpha_{1,ij} S_{ij}(t) + \alpha_{2,ij} \max(S_{ij}(t) - \gamma_{2,ij}, 0) + \alpha_{3,ij} \max(\gamma_{3,ij} - S_{ij}(t), 0)$$

where $\alpha_{n,ij}$ are the slopes and $\gamma_{n,ij}$ are the hinge function knots, visible as breakpoints in the prediction. The MARS terms are selected and pruned using a forward and backward pass. Implemented with the Earth package in R [3], an R-squared threshold of 0.01 limited the addition of terms in the forward pass. The backward pass used generalized cross validation (GCV) as a form of regularization to choose the best subset of terms [4]. The residuals were used to calculate the standard error in each coefficient, and the relative magnitude of the slope compared with the error used to judge whether the slope significantly deviated from zero during each period. We defined a plateau as occurring where the slope of the prediction was not significantly different from zero. One type of plateau we observed was a delayed impact, where a new CI period would initially have zero slope before starting to affect change in the electricity demand. The other type of plateau we observed indicated saturation, where a section with zero slope occurred after the start of the CI period. For the special case where the only slope through a CI period was zero, we considered that saturation as well, since it showed that additional days at that CI had no additional impact.

**Statistical Analysis of Forecasting Accuracy.** Week-ahead demand forecast data and actual demand data from 27 European countries and California (where week-ahead forecast data was available) was used to evaluate the load prediction performance of system operators during the COVID-19 period. The time resolution of the week-ahead forecast is at the hourly level for California and at the daily level (min/max daily demand) for all European countries. Using the week-ahead forecast and actual demand for each country or region for each week, we applied a t-test to the difference to test if the forecast errors were statistically different from 0 based on the p-value obtained. If the difference was statistically significant, we further categorized whether the system operator had over- or under-predicted for that week (depending on the sign of the t-statistic). The threshold of the p-value for significance was set to 0.2 due to the limited number of measurements in European countries (only 14 per week). These results are shown in **Fig. S7**. Data shows that day-ahead forecasts (as opposed to week-ahead forecasts) were corrected within several days.

**Clustering.** K-Means clustering was used to segment the regions based on electricity demand impacts. Use of the Euclidean distance metric allows these clusters to capture similarities in both the timing and magnitude of the response. To select the number of clusters we consulted the elbow curve of K-Means cluster inertia shown in **Fig. S2**. The groupings were robust to the random initialization of the K-Means algorithm.

**Extended Results**

**Relationship to Government Restrictions.** The OLS regression of daily change in electricity demand against CI level presented in **Fig. 2** was run for each country separately. To consider the effect of cluster groupings on the coefficient values, the mean over countries in each cluster is presented in **Fig. S3**. The range of values in each cluster is presented in **Table S3a**. Details on the R-squared values for these regressions are included in **Table S4a**. For the MARS regression, further results are presented in **Fig. S4** and R-squared values are included in **Table S4b**. We consider saturation specifically in **Fig. S4a**, where saturation is defined by a plateau which occurs after the first slope in the period. We use the first slope in each CI period in **Fig. S4b** to study the initial impact of the new CI level. In CI levels 2 and 3 this captures the drop before saturation. We show the timing of the breakpoints in each CI period in **Fig. S4c**. Finally, the value achieved by the end of a period shows the cumulative effect of all CI days leading up to that date, as shown in **Fig. S4d**.

**Relationship to Changes in Mobility.** The elasticity of daily change in electricity demand with respect to each of the six mobility metrics is presented in **Fig. S5**: residential, workplace, retail and recreation, grocery and pharmacy, transit stations, and parks. The range of these values within each cluster is presented in **Table S3b**, and the corresponding $R^2$ results are presented in **Table S4c**.

**Changes in Daily Electricity Demand Patterns.** To further investigate the change in load shapes, in **Fig. S6** we compare the Euclidian distances 1) between April 2020 workday and historic April weekend load curves and 2) between April 2020 workday and historic April workday load curves. Grid operators' forecasting error during this period is studied in **Fig. S7**.

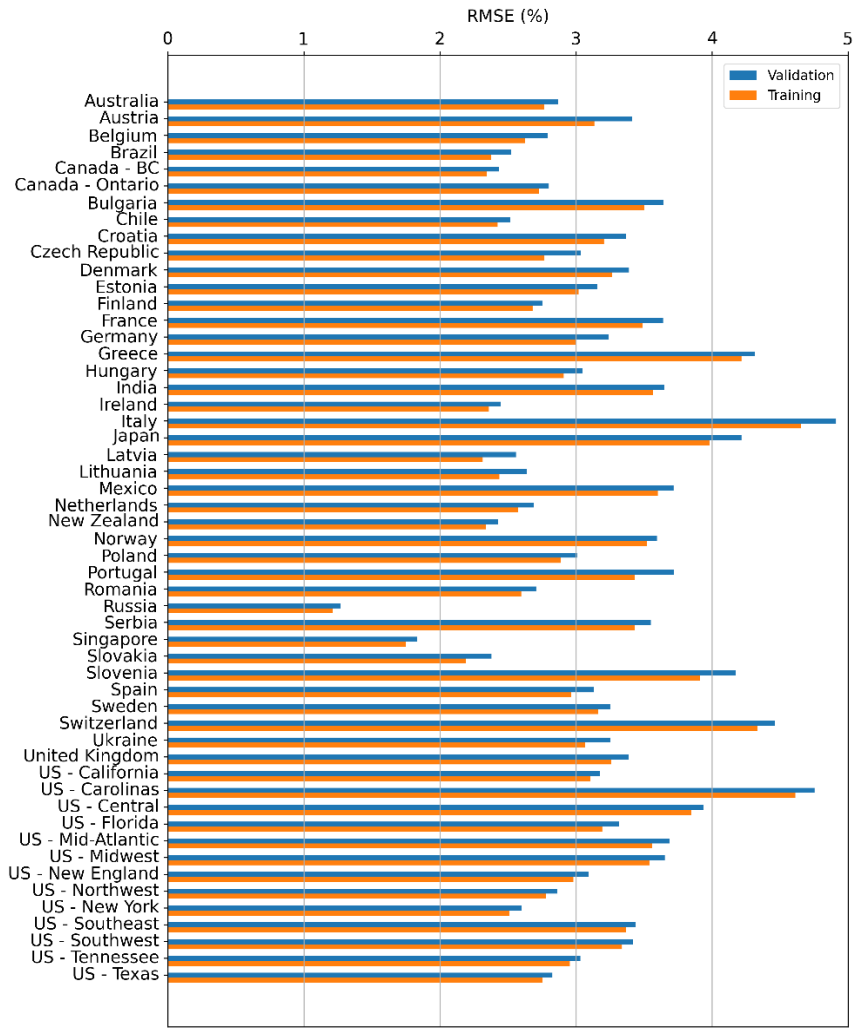

**Fig. S1. Mean in-sample training errors and out-of-sample validation errors for the demand regression model for predicting the daily demand of each region.** Values represent the mean RMSE over all 10 folds of the 10-fold cross validation and are expressed as a percent of the mean demand for each region.

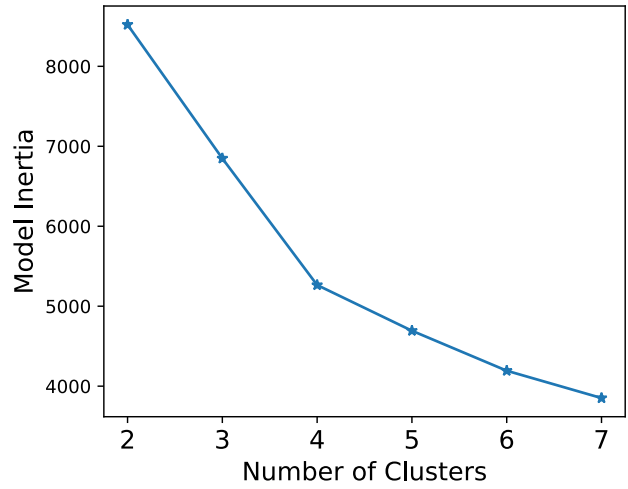

**Fig. S2. Elbow curve for K-means clustering of electricity demand impact.** For each value of K, the model inertia is the sum of the squared distances between the samples and their closest cluster center. This curve shows a kink at K=4.

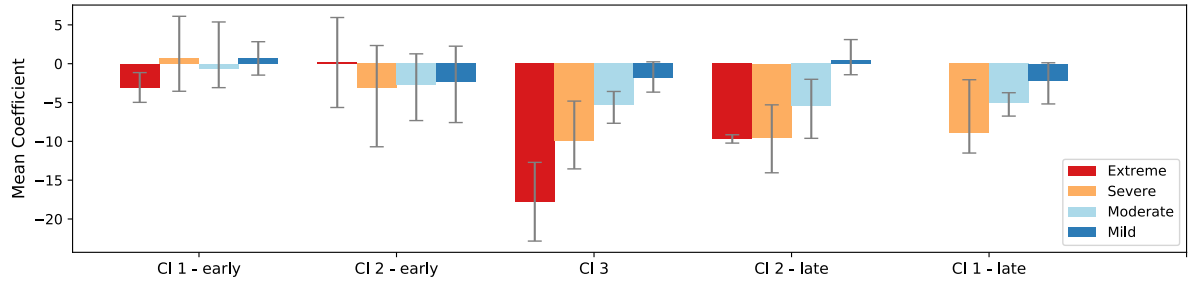

**Fig. S3. Mean coefficients by demand impact group for three different regressions of government restrictions.** Indicators for CI levels 1, 2, and 3 separating the early period of tightening restrictions from the late period of loosening restrictions. All four clusters are shown in red (Extreme), orange (Severe), light blue (Moderate), and dark blue (Mild). The error bars indicate the range of values across all countries in the cluster, and the bar presents the mean. Countries which did not experience a particular level are excluded from the regression.

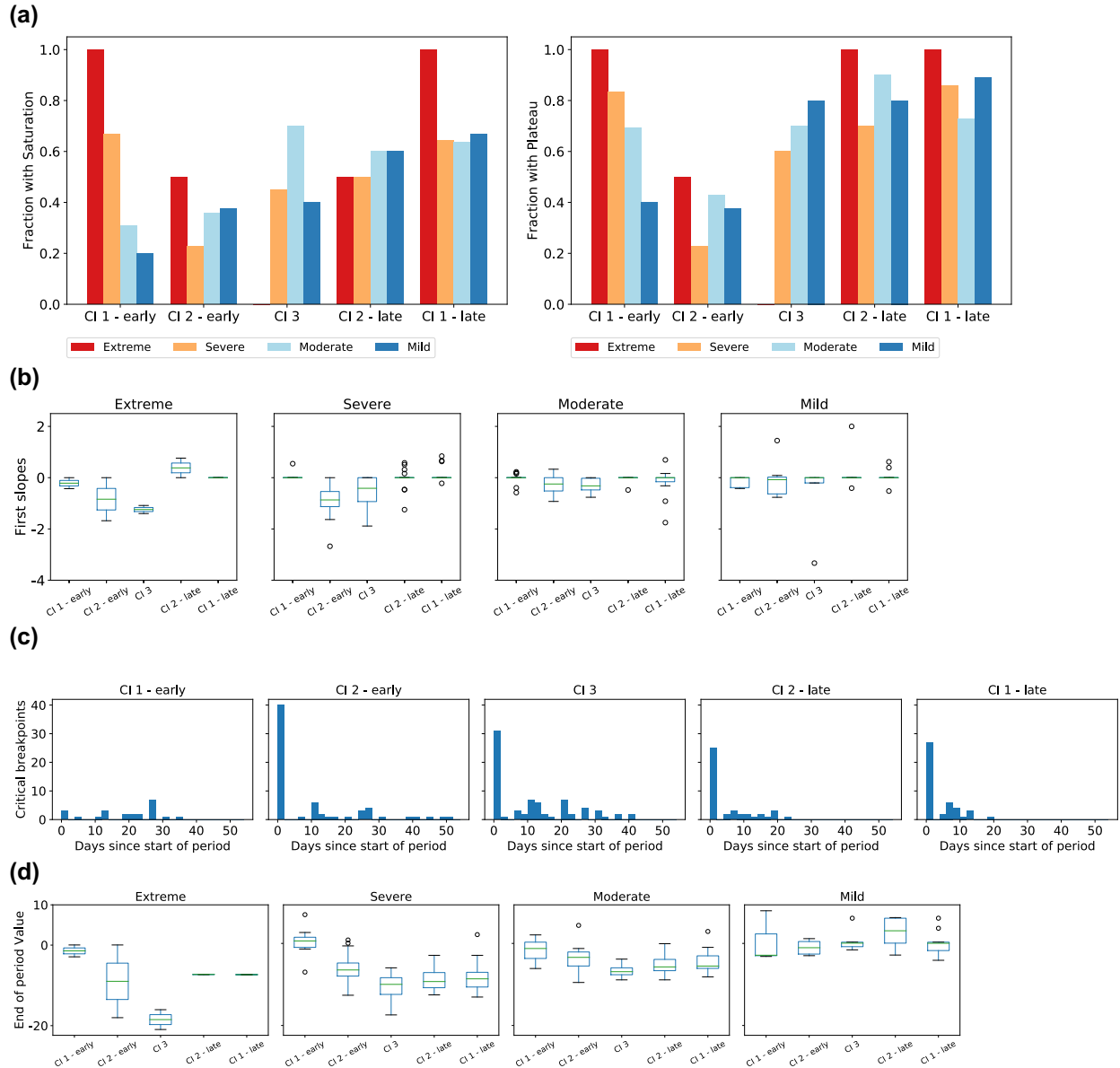

**Fig. S4. MARS regression plots of daily changes in electricity demand with CI level. (a)** Fraction of countries in a cluster which experience a plateau within each CI level (right), or a saturation (left). A plateau is defined as a segment with slope not significantly different from 0, which indicates that each with additional day under the CI level during that segment it is not impacting the demand. Saturation is defined as a plateau which occurs after the first slope of the period, if there are multiple slopes. Each fraction includes only countries which spend time at that CI. **(b)** The first slope of the MARS fit within each CI period, for each country in the cluster. The box covers the middle two quartiles of the data with a line at the median. The whiskers extend to the smaller of the range of the data, or the farthest point within 1.5 times the box height. Datapoints beyond that distance from the box are presented as separate dots. In the Severe cluster in the CI 2 – early period, one outlier out of the range of the plot occurs for Austria at -12.4. **(c)** Histograms showing all significant breakpoints which occur in each CI period, defined as a point where the slope of the MARS prediction changed by greater than 0.1. **(d)** The value achieved at the end of each CI period by the MARS model fit, for countries in each cluster. The box plot boxes, lines, and whiskers are defined as in (b).

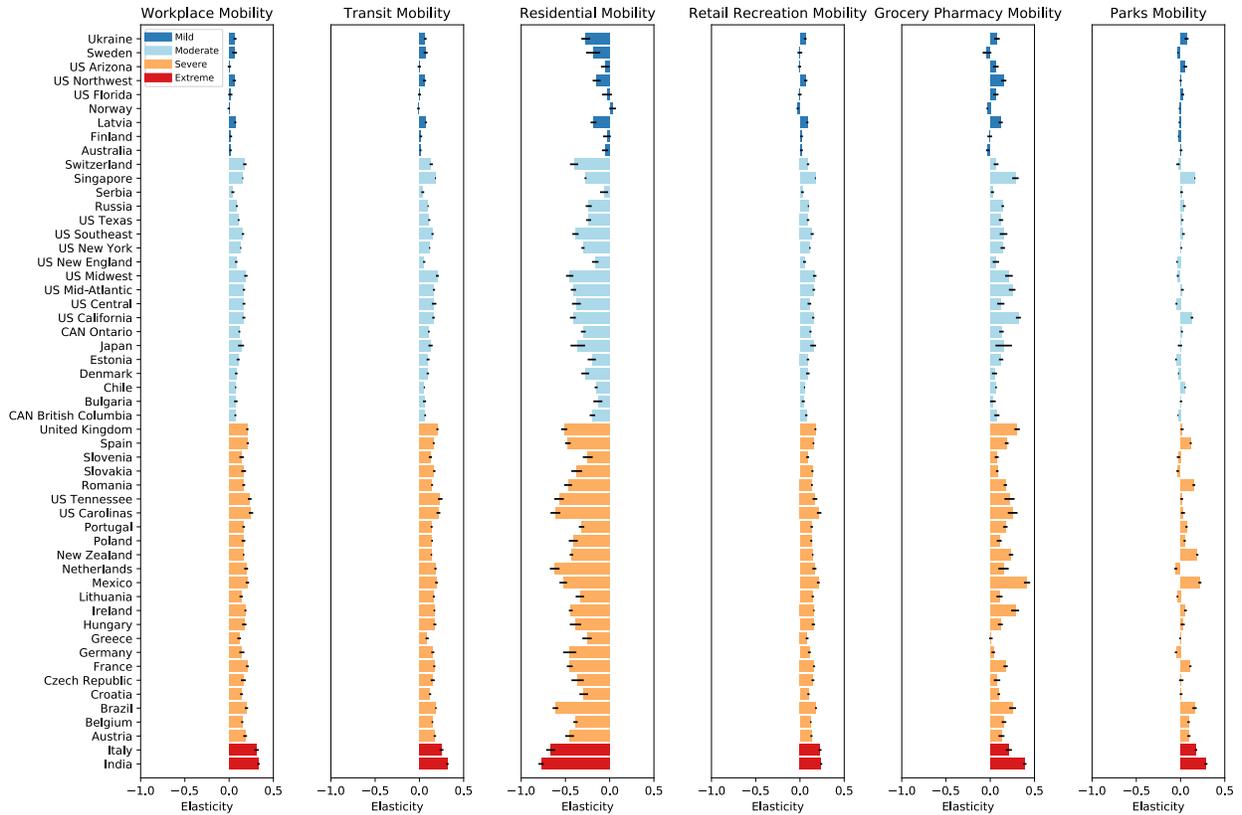

**Fig. S5. Coefficients for mobility metrics from the elasticity calculation (Equation 3).** Bars depict estimates from regressed percent change in daily electricity demand from February 15 to May 25, 2020 against each mobility metric. Colors represent impact groups. All coefficients are significant at $p<0.05$ except: Australia (all); Greece, Japan, and Bulgaria, (Grocery and Parks); Germany, Switzerland, Latvia, Czechia, Hungary, US Carolinas, Slovenia, United Kingdom, Croatia, CAN Ontario, US Mid-Atlantic, US New York, US Tennessee, US Texas, and US Northwest (Parks); Denmark and US New England (Grocery); Finland (Workplaces, Retail, Grocery, Transit, and Residential); Norway (Workplaces, Transit, and Residential); US Florida (Retail, Parks, Transit, Workplaces, and Residential); US Southwest (Retail, Transit, Workplaces, and Residential); Serbia (Retail, Grocery, Parks, and Residential); Sweden (Retail and Grocery).

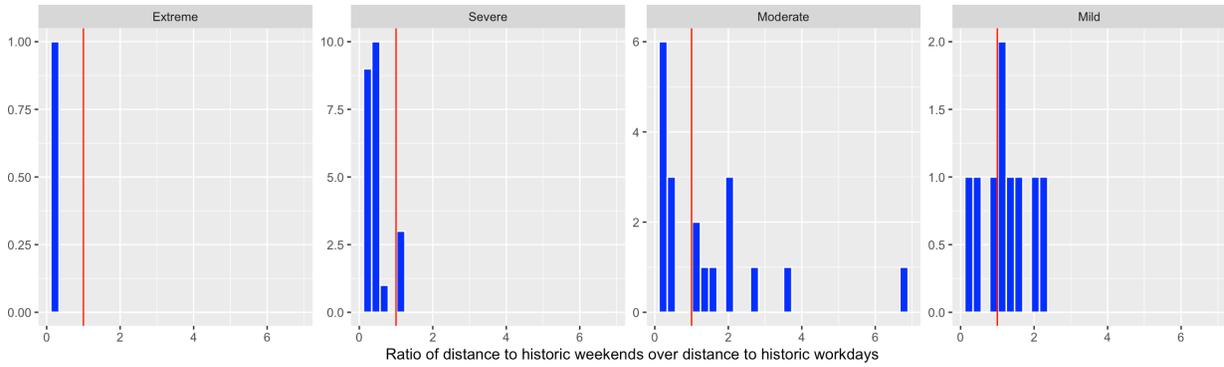

**Fig. S6. Histograms of the ratio of the distance between April 2020 workday and historic April weekend load curves to the distance between April 2020 workday and historic April workday load curves, by impact group.** Distances are computed as the L2 distance between the load curve of median hourly load levels in April 2020 and the load curve of median hourly load levels for the historic period of interest (April weekends in 2016 to 2019 for the numerator and April workdays in 2016 to 2019 for the denominator). Ratios lower than 1 indicate a closer proximity to historic weekends than to historic workdays.

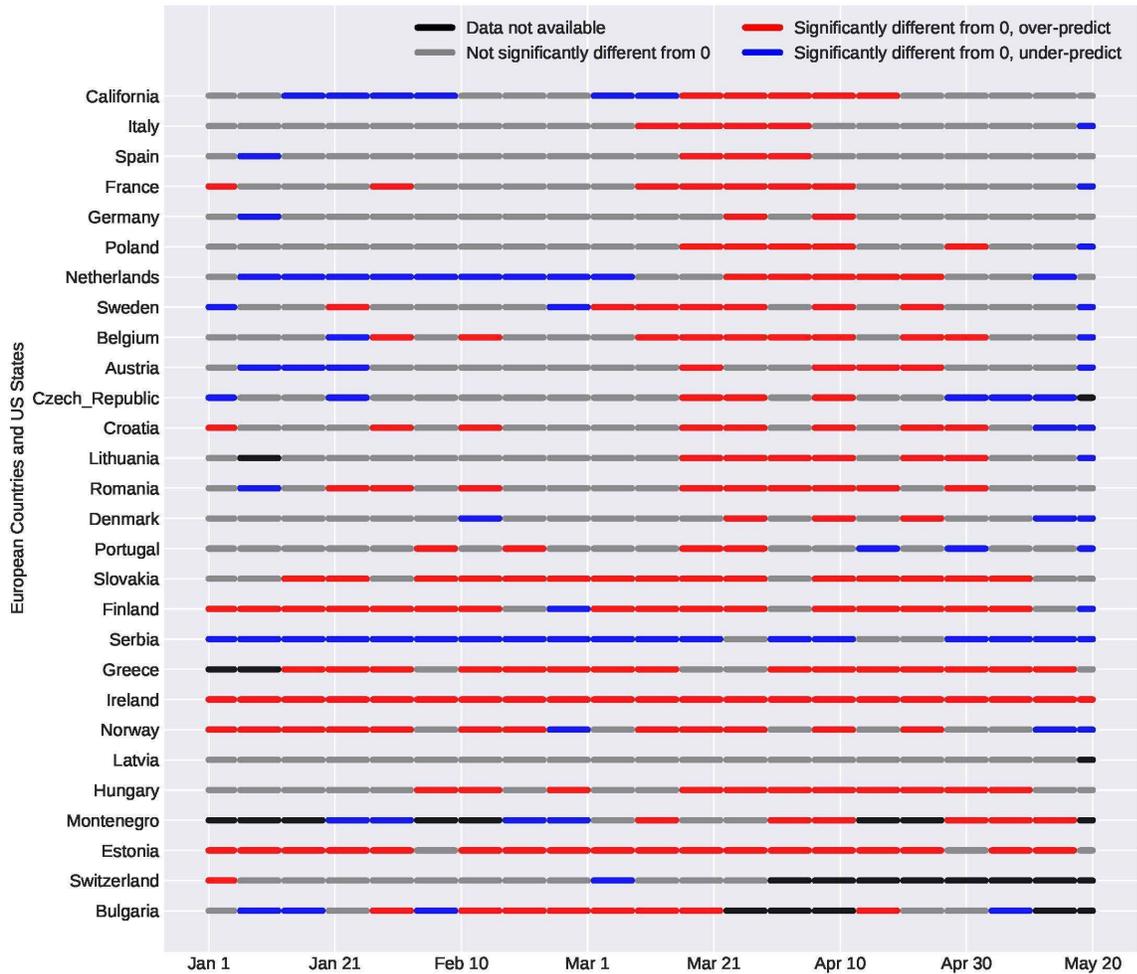

**Fig. S7. Weekly demand forecast accuracy.** Gray line segments represent accurate week-ahead forecasts (not significantly different from actual demand, p<0.20); red over-predictions; and blue under-predictions from Jan 1 to May 20, 2020. Black lines represent missing data.

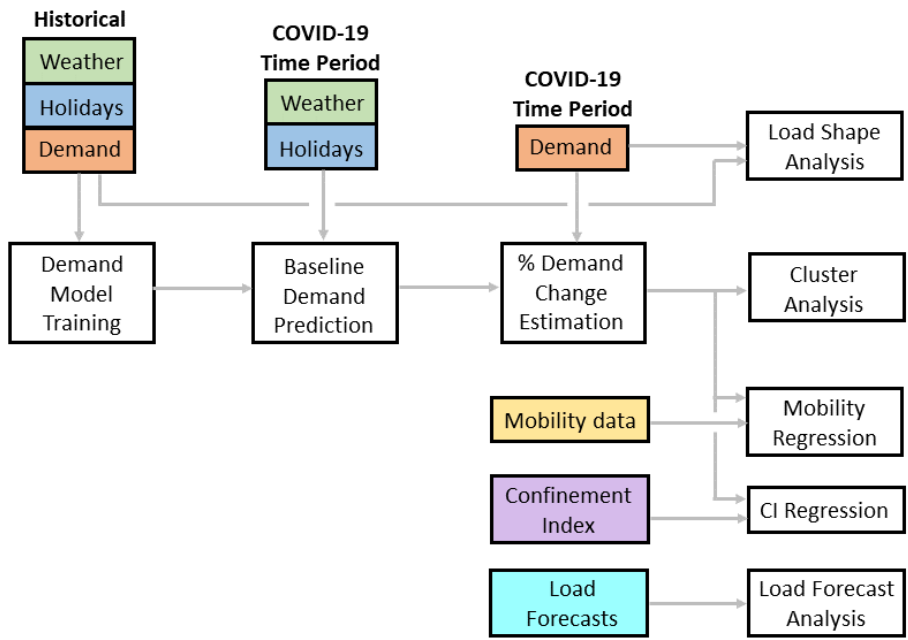

**Fig. S8.** Diagram depicting overall analytic framework

**Table S1.** Data information.

| Coverage / description | Data type | Source |
|---|---|---|
| [D1] United States sub-regional (13 regions) | Electricity, hourly interval | https://www.eia.gov/beta/electricity/gridmonitor/dashboard/electric_overview/US48/US48 |
| [D2] European country-level (29 countries) | Electricity, hourly/ sub-hourly interval | https://transparency.entsoe.eu/ |
| [D3] Australia (5 states) | Electricity, hourly interval | https://www.aemo.com.au/energy-systems/electricity/national-electricity-market-nem/data-nem/aggregated-data |
| [D4] Ontario, Canada | Electricity, hourly interval | http://www.ieso.ca/en/power-data/data-directory |
| [D5] British Columbia | Electricity, hourly interval | https://www.bchydro.com/energy-in-bc/operations/transmission/transmission-system/balancing-authority-load-data.html |
| [D6] Japan (TEPCO utility only) | Electricity, hourly interval | https://www.tepco.co.jp/en/forecast/html/download-e.html |
| [D7] New Zealand | Electricity, hourly interval | https://www.emi.ea.govt.nz/Wholesale/Reports/W_GD_C?DateFrom=20100101&DateTo=20161231&_si=_dr_DateFrom%7C20150101,_dr_DateTo%7C20151231,_dr_RegionType%7CNZ,v%7C4 |
| [D8] Russia | Electricity, hourly interval | http://www.so-cdu.ru/index.php?id=972&tx_ms1cdu_pi1 |
| [D9] Mexico | Electricity, hourly interval | http://www.ons.org.br/Paginas/resultados-da-operacao/historico-da-operacao/curva_carga_horaria.aspx |
| [D10] Brazil | Electricity, hourly interval | http://www.ons.org.br/Paginas/resultados-da-operacao/historico-da-operacao/curva_carga_horaria.aspx |
| [D11] Chile | Electricity, hourly interval | https://www.coordinador.cl/operacion/graficos/demanda/demanda-real-demanda/ |
| [D12] India | Electricity, daily interval | https://posoco.in/reports/daily-reports/daily-reports-2020-21/ |
| [D13] Singapore | Electricity, 30-minute interval | https://www.ema.gov.sg/TemStatistic.aspx?pagesid=20140926wbNYp2Yh8iqy&pagemode=live&sta_sid=20140826Y84sgBebjwKV |
| [D14] China | Electricity, monthly interval | http://www.cec.org.cn/menu/index.html?541 |
| [D15] South Africa | Electricity, monthly interval | https://southafrica.opendataforafrica.org/wsblplg/electricity-generated-and-available-for-distribution-of-south-africa-monthly-update |
| [D16] Thailand | Electricity, monthly interval | http://www.eppo.go.th/index.php/en/en-energystatistics/electricity-statistic |
| [D17] Argentina | Electricity, monthly interval | https://portalweb.cammesa.com/Memnet1/default.aspx |
| [D18] Kenya | Electricity, monthly interval | https://www.knbs.or.ke/?wpdmpro=leading-economic-indicators-may-2020 |
| [D19] NOAA climate data (US) | Degree days | https://www.cpc.ncep.noaa.gov/products/analysis_monitoring/cdus/degree_days/ |
| [D20] ASOS weather system network (global) | Temperature, daily | https://mesonet.agron.iastate.edu/request/download.phtml |
| [D21] Public holidays and observances (global) | Selected dates | https://www.timeanddate.com/holidays/ |
| [D22] Google Community Mobility Report: changes in human mobility patterns (global) | % change from baseline, daily | https://www.google.com/covid19/mobility/ |
| [D23] COVID-19 Confinement Index (global), Le Quéré et al. (2020) | Daily | https://www.icos-cp.eu/gcp-covid19<br>https://www.nature.com/articles/s41558-020-0797-x |

**Table S2.** Definition of the CI levels from [1] available at [2].

| Level | Description | Policy examples |
|---|---|---|
| 0 | No restrictions | |
| 1 | Policies targeted at long distance travel or groups of individuals where outbreak first nucleates | Isolation of sick or symptomatic individuals |
| | | Self-quarantine of travelers arriving from affected countries |
| | | Screening passengers at transport hubs |
| | | Ban of mass gatherings >5,000 |
| | | Closure of selected national borders and restricted international travel |
| | | Citizen repatriation |
| 2 | Regional policies that restrict an entire city, region or ~50% of society from normal daily routines | Closure of all national borders |
| | | Mandatory closure of schools, universities, public buildings, religious or cultural buildings, restaurants, bars and other non-essential businesses within a city or region |
| | | Ban of public gatherings >100 |
| | | Perhaps also accompanied by recommended closures at a broader or national level |
| | | Mandatory night curfew |
| 3 | National policies that substantially restrict the daily routine of all but key workers | Mandatory national 'lockdown' that requires household confinement of all but key workers |
| | | Ban public gatherings and enforce social distancing >2 m |

**Table S3.** Coefficients for ordinary least square regression models predicting change in electricity demand for individual countries/regions with confinement index indicators (A) and changes in workplace mobility (B). Both regressions are restricted to the period 2/15/20-5/31/20.

(A)

| Independent Variable | | Cluster | | | | All Countries |
|---|---|---|---|---|---|---|
| | | Extreme | Severe | Moderate | Mild | |
| CI 1 - early | Mean | -3.063 | 0.678 | -0.634 | 0.659 | -0.092 |
| | Min | -4.978 | -3.549 | -3.085 | -1.468 | -4.978 |
| | Max | -1.148 | 6.105 | 5.375 | 2.832 | 6.105 |
| CI 2 - early | Mean | 0.154 | -3.096 | -2.759 | -2.334 | -2.720 |
| | Min | -5.640 | -10.699 | -7.325 | -7.581 | -10.699 |
| | Max | 5.948 | 2.341 | 1.270 | 2.256 | 5.948 |
| CI 3 | Mean | -17.776 | -9.928 | -5.341 | -1.828 | -8.018 |
| | Min | -22.845 | -13.548 | -7.672 | -3.662 | -22.845 |
| | Max | -12.707 | -4.818 | -3.576 | 0.242 | 0.242 |
| CI 2 - late | Mean | -9.688 | -9.500 | -5.447 | 0.512 | -7.062 |
| | Min | -10.225 | -14.049 | -9.612 | -1.422 | -14.049 |
| | Max | -9.151 | -5.298 | -2.010 | 3.106 | 3.106 |
| CI 1 - late | Mean | - | -8.914 | -5.025 | -2.187 | -6.260 |
| | Min | - | -11.507 | -6.748 | -5.182 | -11.507 |
| | Max | - | -2.064 | -3.738 | 0.120 | 0.120 |

(B)

| Independent Variable (Mobility) | | Cluster | | | | All Countries |
|---|---|---|---|---|---|---|
| | | Extreme | Severe | Moderate | Mild | |
| Workplace | Mean | 0.322 | 0.176 | 0.120 | 0.034 | 0.137 |
| | Min | 0.310 | 0.113 | 0.042 | -0.004 | -0.004 |
| | Max | 0.334 | 0.247 | 0.190 | 0.070 | 0.334 |
| Transit | Mean | 0.283 | 0.163 | 0.115 | 0.032 | 0.128 |
| | Min | 0.250 | 0.088 | 0.038 | -0.010 | -0.010 |
| | Max | 0.317 | 0.236 | 0.203 | 0.074 | 0.317 |
| Residential | Mean | -0.721 | -0.435 | -0.281 | -0.102 | -0.334 |
| | Min | -0.775 | -0.621 | -0.452 | -0.270 | -0.775 |
| | Max | -0.667 | -0.248 | -0.064 | 0.034 | 0.034 |
| Retail and Recreation | Mean | 0.233 | 0.145 | 0.104 | 0.023 | 0.113 |
| | Min | 0.228 | 0.081 | 0.028 | -0.026 | -0.026 |
| | Max | 0.239 | 0.217 | 0.175 | 0.083 | 0.239 |
| Grocery and Pharmacy | Mean | 0.301 | 0.164 | 0.132 | 0.041 | 0.137 |
| | Min | 0.212 | 0.007 | 0.024 | -0.037 | -0.037 |
| | Max | 0.391 | 0.416 | 0.318 | 0.153 | 0.416 |
| Parks | Mean | 0.231 | 0.052 | 0.013 | 0.008 | 0.037 |
| | Min | 0.175 | -0.053 | -0.051 | -0.030 | -0.053 |
| | Max | 0.287 | 0.218 | 0.162 | 0.068 | 0.287 |

**Table S4.** R-squared results for the ordinary least squares regressions presented in Table S2. (A) for regression with confinement indices, (b) for MARS model, (c) for mobility elasticity regressions

(a)

|  | Cluster | | | | All Countries |
|---|---|---|---|---|---|
|  | Extreme | Severe | Moderate | Mild |  |
| Min R2 | 0.584 | 0.195 | 0.043 | 0.047 | 0.043 |
| Max R2 | 0.824 | 0.855 | 0.806 | 0.388 | 0.855 |

(b)

|  | Cluster | | | | All Countries |
|---|---|---|---|---|---|
|  | Extreme | Severe | Moderate | Mild |  |
| Min R2 | 0.850 | 0.648 | 0.357 | 0.250 | 0.250 |
| Max R2 | 0.905 | 0.908 | 0.840 | 0.665 | 0.908 |

(c)

| Independent Variable (Mobility) | | Cluster | | | | All Countries |
|---|---|---|---|---|---|---|
| | | Extreme | Severe | Moderate | Mild | |
| Workplace | Min | 0.716 | 0.230 | 0.066 | 0.000 | 0.000 |
|  | Max | 0.795 | 0.804 | 0.895 | 0.334 | 0.895 |
| Transit | Min | 0.615 | 0.191 | 0.075 | 0.000 | 0.000 |
|  | Max | 0.803 | 0.760 | 0.902 | 0.306 | 0.902 |
| Residential | Min | 0.732 | 0.147 | 0.021 | 0.003 | 0.003 |
|  | Max | 0.804 | 0.788 | 0.880 | 0.255 | 0.880 |
| Retail Recreation | Min | 0.651 | 0.191 | 0.039 | 0.000 | 0.000 |
|  | Max | 0.673 | 0.804 | 0.900 | 0.295 | 0.900 |
| Grocery Pharmacy | Min | 0.362 | 0.001 | 0.009 | 0.000 | 0.000 |
|  | Max | 0.855 | 0.600 | 0.560 | 0.253 | 0.855 |
| Parks | Min | 0.579 | 0.001 | 0.001 | 0.000 | 0.000 |
|  | Max | 0.592 | 0.627 | 0.860 | 0.171 | 0.860 |

**SI References**